\documentclass[12pt]{article}
\usepackage{eurosym}
\usepackage{float}
\usepackage{graphicx}
\usepackage[utf8]{inputenc}
\usepackage[T1]{fontenc}
\usepackage{indentfirst}
\usepackage[margin=0.6in,nomarginpar]{geometry}
\usepackage[final]{hyperref}
\usepackage{amsmath}
\usepackage{hyperref}
\usepackage{cite}
\usepackage{subcaption}
\usepackage{caption}
\usepackage{amssymb}
\usepackage{multirow}
\usepackage[table]{xcolor}
\usepackage{orcidlink}
\usepackage{color}
\usepackage{hyperref}
\usepackage[utf8]{inputenc}
\hypersetup{
colorlinks=true,
linkcolor=blue,
citecolor=blue,
filecolor=magenta,
urlcolor=blue
}

\begin{document}

\title{Spectral and Thermal Analysis of the Morse Potential within the Dunkl Formalism: Analytical Approximations and Applications}

\author{B. Hamil \orcidlink{0000-0002-7043-6104} \thanks{%
hamilbilel@gmail.com} , \\
Laboratoire de Physique Math\'{e}matique et Subatomique,\\
Facult\'{e} des Sciences Exactes, Universit\'{e} Constantine 1, Constantine,
Algeria.  \\ 
B. C. L\"{u}tf\"{u}o\u{g}lu 
\orcidlink{Orcid ID :
0000-0001-6467-5005} \thanks{%
bekir.lutfuoglu@uhk.cz (Corresponding author)} , \\
Department of Physics, Faculty of Science, University of Hradec Kralove, \\
Rokitanskeho 62/26, Hradec Kralove, 500 03, Czech Republic. \\
A. N. Ikot \orcidlink{0000-0002-1078-262X} \thanks{%
ndemikotphysics@gmail.com} , \\
Department of Physics, Theoretical Physics Group, \\ University of Port Harcourt, Choba, Nigeria. \and
U. S. Okorie \orcidlink{0000-0002-5660-0289} \thanks{%
okoriu@unisa.ac.za} , \\
Department of Physics, University of South Africa, \\
Florida 1710, Johannesburg, South Africa.}
\date{\today }
\maketitle

\begin{abstract}
In this work, we investigate the quantum dynamics of a particle subject to the Morse potential within the framework of Dunkl quantum mechanics. By employing the Dunkl derivative operator—which introduces reflection symmetry—we construct a deformed Schrödinger equation and obtain exact analytical solutions using the Pekeris approximation. The resulting energy spectrum and wavefunctions reveal how Dunkl parameters alter the effective potential and vibrational states. The model is applied to several diatomic molecules, including H$_2$, HCl, and I$_2$, illustrating the impact of symmetry deformation on energy spectra. We also compute thermodynamic functions including the partition function, free energy, internal energy, entropy, and specific heat. The analysis shows that the Dunkl deformation induces distinct thermal behavior and offers a tunable approach to molecular modeling. These results highlight the potential of the Dunkl formalism as a useful tool for extending conventional quantum models and for exploring symmetry-deformed systems in molecular physics and quantum thermodynamics.
\end{abstract}

\section{Introduction}

The Morse potential, first introduced by Philip M. Morse in 1929~\cite{Morse1929}, is one of the most widely used anharmonic potential functions in theoretical chemistry and physics. It is mathematically expressed as
\begin{equation}
 V(r) = D \left(1 - e^{-\alpha(r-r_e)}\right)^2, \label{eq1}
\end{equation}
where $D$ is the dissociation energy, {$r$ is the distance from the origin, }$r_e$ is the equilibrium bond distance, and $\alpha$ determines the potential well's width. Unlike the harmonic oscillator, which assumes a perfectly quadratic potential, the Morse potential accounts for bond anharmonicity, providing a more realistic description of molecular vibrations and dissociation processes~\cite{Levine2009}. This feature makes it particularly valuable in spectroscopy, enabling accurate predictions of vibrational spectra, including high-energy transitions observed in infrared and Raman spectroscopy~\cite{Wilson1955}. Furthermore, its parameters correspond directly to measurable molecular properties, facilitating widespread application in computational chemistry, including molecular dynamics simulations~\cite{Leach2001}, density functional theory calculations~\cite{Parr1989}, and semi-empirical models~\cite{Thiel2014}. However, despite its success in modeling diatomic interactions and localized bonds in larger molecules, the Morse potential’s applicability to more complex structures remains limited without suitable extensions~\cite{Herzberg1950}.

Beyond molecular modeling, the Morse potential plays a foundational role in various fields of physics~\cite{Mirzanejad2024}. In quantum mechanics, it serves as an exactly solvable model~\cite{Flugge1999, Znojil2016}, providing a benchmark for testing approximation methods~\cite{Dong2007}. In nuclear physics~\cite{Greiner1995} and statistical mechanics~\cite{McQuarrie1986}, it helps describe effective interatomic interactions. In condensed matter and solid-state physics, it is used to study nonlinear lattice vibrations and phonon dynamics~\cite{Allen1989, Dove1993, Ibach2009}. It has also found use in atomic physics to model Rydberg molecule interactions~\cite{Gallagher1994} and interatomic forces in crystal structure analysis. However, the exponential decay of the Morse potential limits its capacity to model long-range interactions, lacking the proper asymptotic behavior for Coulombic and van der Waals forces~\cite{Stone2013}. To overcome this, several generalized and hybrid extensions of the Morse potential have been proposed~\cite{Yang2018}. Its utility extends even further into interdisciplinary research in biophysics, materials science, and engineering, where modified forms of the Morse potential are used to study protein-ligand interactions, DNA elasticity~\cite{Zdravkovic2006}, and force spectroscopy experiments~\cite{Zdravkovic2012}, yielding insights into molecular flexibility and stability~\cite{Ayyappan2022}.

Conversely, Dunkl operator theory offers a robust mathematical framework for analyzing systems in which symmetry—particularly parity symmetry—plays a central role in governing analytical solutions and underlying structure. By extending classical differential operators with reflection operators, Dunkl operators encode both continuous and discrete symmetries. Introduced by Charles F. Dunkl in the late 1980s~\cite{Dunkl1989}, the formalism was initially developed to explore special functions and orthogonal polynomials related to reflection groups~\cite{Dunkl1988}. Since then, Dunkl operators have found diverse applications in mathematics and physics, particularly in harmonic analysis~\cite{Heckman1991}, multivariate orthogonal polynomials~\cite{Dunkl2001}, and integrable systems with reflection symmetry~\cite{Rosler2004}. Physically, they are integral to studies of deformed Heisenberg algebras~\cite{Plyushchay1994, Plyushchay1996, Miky1}, root system symmetries~\cite{Hikami1996}, and quantum systems exhibiting reflection invariance~\cite{Plyushchay1997, Miky2, Miky3}. The Dunkl framework also supports research into bosonized supersymmetry~\cite{Gamboa1999, Plyushchay2000, Horvathy2010}, fractional helicity~\cite{Klishevich2001}, anyonic systems~\cite{Horvathy2004}, Calogero–Sutherland models~\cite{Kakei1996, Lapointe1996}, and Wigner–Heisenberg algebra realizations in atomic physics~\cite{Rodrigues2009}.

In recent years, Dunkl operator methods have demonstrated exceptional versatility in tackling quantum systems with reflection symmetry. They have yielded exact solutions in quantum many-body models~\cite{Genest20131, Genest20132, Genest20141, Isaac2016, Ghazouani2020, Najafizade20223} and offered new perspectives on condensed matter systems~\cite{Bilel5}. Their success spans both non-relativistic~\cite{Genest20142, Salazar2017, Mota2022s, Halberg2022, Samira20222, Dong2023} and relativistic domains~\cite{Sargol2018, Mota20181, Ojeda2020, Mota20211, Mota20212, Bilel20221, Merad2021, Merad2022, Askari2023}, including thermodynamic analysis~\cite{Ubriaco2014, Dong2021, Bilel20222, Rouabhia2023}. Recent developments have expanded the formalism to include time-dependent dynamics~\cite{Benchikha20241, Lutfuoglu2025}, Coulomb potentials~\cite{Vincent3, Salazar2018, Ghazouani2019, Ghazouani2021}, noncommutative geometries~\cite{Samira2022}, and curved spacetime settings~\cite{Najafizade20221, Najafizade20222, Ballesteros2023, Askour2025}. The Dunkl framework continues to yield insights into statistical mechanics~\cite{HH2021, phys5, Bilel3, HO24, phys6}, quantum information theory~\cite{Debraj2024}, and generalized quantum models~\cite{Junker2023, Quesne2024, Bougerne, Bouguerne20243, Junker20241, Benzair2024, Benchikha20242, Schulze20243, Schulze20244, Schulze20245, Schulze20246, Mota20241, Mota20242, Hassanabadi2024, Benzair2025, Hamil20251, Benchikha2025, Ham20251}, attesting to its wide applicability in modern physics.

The Morse potential’s enduring role in molecular and quantum physics, coupled with the evolving Dunkl operator formalism, presents a compelling platform for exploring quantum dynamics through reflection symmetry. This study unites these perspectives by examining a Morse-type system within the Dunkl framework. Our approach yields analytical solutions that demonstrate how symmetry deformation influences anharmonic behavior while preserving essential physical characteristics. These findings contribute new insight to both theoretical chemistry—where subtle spectral deviations may indicate novel quantum effects—and mathematical physics, where the Morse potential serves as an ideal testbed for symmetry-deformed quantum mechanics. By providing exact solutions and thermodynamic analyses, we establish a foundation for further exploration of reflection-symmetric quantum systems, particularly in molecular spectroscopy and related fields.

This paper is organized as follows. In Section~\ref{sec2}, we reformulate the Schrödinger equation in the presence of the Morse potential using the Dunkl formalism, introducing a potential form suitable for exact treatment. Section~\ref{sec3} presents the analytical solution of the radial equation, where we obtain closed-form expressions for the energy eigenvalues and wavefunctions. Section~\ref{sec4} is devoted to the spectral analysis and molecular applications of the model, where the theoretical predictions are compared with experimental data for diatomic molecules such as H$_2$, HCl, and I$_2$. This section highlights how the Dunkl deformation enriches the vibrational structure and leads to modified level spacings. In Section~\ref{sec5}, we investigate the thermal behavior of the system, deriving key thermodynamic quantities—partition function, internal energy, entropy, and specific heat—from the exact spectrum. The influence of temperature and deformation parameters is examined in detail, demonstrating the broader applicability of the model to quantum statistical systems. Finally, Section~\ref{sec6} summarizes our findings and outlines potential extensions, particularly in the context of deformed quantum mechanics and molecular thermodynamics.



\section{The Dunkl-Schrödinger Equation in Spherical Coordinates} \label{sec2}

In this section, we formulate the time-independent Dunkl-Schrödinger equation for a particle in a spherically symmetric potential. We start by introducing the Dunkl operators and their associated reflection symmetries, then proceed to express the Dunkl-Laplacian in spherical coordinates. Finally, we separate variables to obtain the angular and radial components, which will serve as the foundation for solving the system in the presence of a Morse potential in the proceeding section.

\subsection{Dunkl Operators and Reflection Symmetries}

In Dunkl quantum mechanics, the conventional derivative in the momentum operator is replaced by the Dunkl derivative $D_j$, leading to a deformed momentum operator of the form
\begin{equation}
p_j = \frac{\hbar}{i} D_j = \frac{\hbar}{i} \left( \partial_{x_j} + \frac{\mu_j}{x_j} (1 - R_j) \right), \qquad j = 1, 2, 3,
\end{equation}
where $\mu_j$ are real deformation parameters and $R_j$ denote the reflection operators. These reflections act on {a function} $f(x_j)$ defined on the real line via
\begin{equation}
R_j f(x_j) = f(-x_j).
\end{equation}
{To ensure square integrability in three dimensions, the deformation parameters must satisfy \( \mu_i > -\frac{1}{2} \). The corresponding weighted norm is
\begin{equation}
\int_{\mathbb{R}^3} |f(x,y,z)|^2\, |x|^{2\mu_1} |y|^{2\mu_2} |z|^{2\mu_3} \, dx\,dy\,dz < \infty.
\end{equation} }
To exploit spherical symmetry, we {then} introduce the standard transformation to spherical coordinates:
\begin{equation}
x_1 = r \sin\theta \cos\varphi, \qquad x_2 = r \sin\theta \sin\varphi, \qquad x_3 = r \cos\theta, 
\end{equation}
{with the domains $\theta \in [0, \pi]$ and $\phi \in [0, 2\pi)$.}

\subsection{Dunkl-Laplacian and Angular Operators}

The squared Dunkl operator, interpreted as a generalized Laplacian, takes the following form in spherical coordinates {\cite{Genest20142}}
\begin{equation}
\Delta^D = D_j^2 = \frac{\partial^2}{\partial r^2} + \frac{2(1 + \mu_1 + \mu_2 + \mu_3)}{r} \frac{\partial}{\partial r} + \frac{\mathcal{J}_\varphi}{r^2 \sin^2\theta} + \frac{\mathcal{J}_\theta}{r^2},
\end{equation}
where $\mathcal{J}_\varphi$ and $\mathcal{J}_\theta$ are the angular Dunkl operators associated with the azimuthal and polar coordinates, respectively. These are given by
\begin{equation}
\mathcal{J}_\varphi = \frac{\partial^2}{\partial \varphi^2} + 2\left[\mu_2 \cot\varphi - \mu_1 \tan\varphi \right] \frac{\partial}{\partial \varphi}
- \frac{\mu_1}{\cos^2\varphi} (1 - R_1) - \frac{\mu_2}{\sin^2\varphi} (1 - R_2),
\end{equation}
and
\begin{equation}
\mathcal{J}_\theta = \frac{\partial^2}{\partial \theta^2} + 2\left[\left( \tfrac{1}{2} + \mu_1 + \mu_2 \right) \cot\theta - \mu_3 \tan\theta \right] \frac{\partial}{\partial \theta}
- \frac{\mu_3}{\cos^2\theta} (1 - R_3).
\end{equation}
The action of the reflection operators in spherical coordinates modifies angular variables as follows:
\begin{equation}
R_1 f(r, \theta, \varphi) = f(r, \theta, \pi - \varphi), \quad
R_2 f(r, \theta, \varphi) = f(r, \theta, -\varphi), \quad
R_3 f(r, \theta, \varphi) = f(r, \pi - \theta, \varphi).
\end{equation}

\subsection{Separation of Variables}

We consider a non-relativistic particle of mass $M$ in a central potential $V(r)$. In standard quantum mechanics, the corresponding time-independent Schrödinger equation reads
\begin{equation}
\left[ \frac{p^2}{2M} + V(r) - E \right] \psi(\vec{r}) = 0.
\end{equation}
Within the Dunkl framework, the momentum operator $p_j$ is replaced by the Dunkl momentum operator, resulting in the modified Schrödinger equation
\begin{equation}
\left[ \frac{\partial^2}{\partial r^2} + \frac{2(1 + \mu_1 + \mu_2 + \mu_3)}{r} \frac{\partial}{\partial r} + \frac{\mathcal{J}_\varphi}{r^2 \sin^2\theta} + \frac{\mathcal{J}_\theta}{r^2}
- \frac{2M}{\hbar^2} V(r) + \frac{2M}{\hbar^2} E \right] \psi(\vec{r}) = 0.
\end{equation}
To solve this equation, we adopt the ansatz
\begin{equation}
\psi(\vec{r}) = r^\delta \Psi(r) \Theta(\theta) \Phi(\varphi),
\end{equation}
which separates the wave function into radial and angular components. {This ansatz is justified by the fact that $r = 0$ is a regular singular point of the radial differential equation, allowing a solution of the form $\psi(r) = r^\delta \Psi(r)$, consistent with the Frobenius method.} Substituting into the Dunkl-Schrödinger equation yields three independent differential equations {\cite{Genest20142}}:
\begin{equation}
\left[ \mathcal{J}_\varphi + \lambda^2 \right] \Phi(\varphi) = 0,
\end{equation}
\begin{equation}
\left[ \mathcal{J}_\theta - \frac{\lambda^2}{\sin^2\theta} + \varpi^2 \right] \Theta(\theta) = 0,
\end{equation}
\begin{equation}
\left[ \frac{d^2}{dr^2} - \frac{\varpi^2 + \delta(\delta + 1)}{r^2} - \frac{2M}{\hbar^2} V(r) + \frac{2M}{\hbar^2} E \right] \Psi(r) = 0, \label{eqrad}
\end{equation}
where $\lambda$ and $\varpi$ are separation constants, and the exponent $\delta$ is given by
\begin{equation}
\delta = -\left(1 + \mu_1 + \mu_2 + \mu_3 \right).
\end{equation}

\subsection{Solutions to the Angular Equations}

We begin with the azimuthal equation. Since the angular operator $\mathcal{J}_\varphi$ commutes with the reflection operators $R_1$ and $R_2$,
\begin{equation}
[\mathcal{J}_\varphi, R_1] = [\mathcal{J}_\varphi, R_2] = 0,
\end{equation}
its eigenfunctions $\Phi(\varphi)$ can be chosen to possess definite parity under these reflections. That is,
\begin{equation}
R_j \Phi(\varphi) = s_j \Phi(\varphi), \quad s_j = \pm 1, \quad j = 1, 2.
\end{equation}
Based on the values of $s_1$ and $s_2$, we identify four distinct classes of solutions:
\begin{itemize}
\item For $s_1 = s_2 = +1$, the solution reads:
\begin{equation}
\Phi_m^{+,+} = \mathcal{C}_m^{+,+} P_m^{(\mu_1 - 1/2, \mu_2 - 1/2)}(-\cos 2\varphi), \quad m \in \mathbb{N},
\end{equation}
where $P_m^{(\alpha, \beta)}(x)$ are Jacobi polynomials.

\item For $s_1 = s_2 = -1$, the solution becomes:
\begin{equation}
\Phi_m^{-,-} = \mathcal{C}_m^{-,-} \sin 2\varphi \, P_{m-1}^{(\mu_1 + 1/2, \mu_2 + 1/2)}(-\cos 2\varphi), \quad m \in \mathbb{N}.
\end{equation}

\item For $s_1 = +1$, $s_2 = -1$, the solution is:
\begin{equation}
\Phi_m^{+,-} = \mathcal{C}_m^{+,-} \sin \varphi \, P_{m - 1/2}^{(\mu_1 - 1/2, \mu_2 + 1/2)}(-\cos 2\varphi), \quad m \in \left\{ \tfrac{1}{2}, \tfrac{3}{2}, \tfrac{5}{2}, \ldots \right\}.
\end{equation}

\item For $s_1 = -1$, $s_2 = +1$, the solution becomes:
\begin{equation}
\Phi_m^{-,+} = \mathcal{C}_m^{-,+} \cos \varphi \, P_{m - 1/2}^{(\mu_1 + 1/2, \mu_2 - 1/2)}(-\cos 2\varphi), \quad m \in \left\{ \tfrac{1}{2}, \tfrac{3}{2}, \tfrac{5}{2}, \ldots \right\}.
\end{equation}
\end{itemize}
In all four cases, the corresponding separation constant $\lambda^2$ takes the unified form:
\begin{equation}
\lambda^2 = 4m(m + \mu_1 + \mu_2).
\end{equation}
It is important to note that when $s_1 s_2 = +1$, $m$ must be a non-negative integer, while for $s_1 s_2 = -1$, $m$ assumes positive half-integer values.

We now turn to the polar equation. Since $\mathcal{J}_\theta$ commutes with $R_3$,
\begin{equation}
[\mathcal{J}_\theta, R_3] = 0,
\end{equation}
we label the corresponding eigenfunctions $\Theta(\theta)$ by the eigenvalue $s_3 = \pm 1$ such that
\begin{equation}
R_3 \Theta(\theta) = s_3 \Theta(\theta).
\end{equation}

\begin{itemize}
\item For $s_3 = +1$, the polar solution is:
\begin{equation}
\Theta_\ell^{+} = \iota_\ell^{+} \sin^{2m}\theta \, P_\ell^{(2m + \mu_1 + \mu_2, \mu_3 - 1/2)}(\cos 2\theta), \quad \ell \in \mathbb{N}.
\end{equation}

\item For $s_3 = -1$, the solution becomes:
\begin{equation}
\Theta_\ell^{-} = \iota_\ell^{-} \sin^{2m}\theta \, P_{\ell - 1/2}^{(2m + \mu_1 + \mu_2, \mu_3 + 1/2)}(\cos 2\theta), \quad \ell \in \left\{ \tfrac{1}{2}, \tfrac{3}{2}, \tfrac{5}{2}, \ldots \right\}.
\end{equation}
\end{itemize}
In both parity cases, the separation constant $\varpi^2$ is given by
\begin{equation}
\varpi^2 = 4(\ell + m)\left(\ell + m + \mu + \tfrac{1}{2}\right), \qquad \mu = \mu_1 + \mu_2 + \mu_3.
\end{equation}

\section{Radial Dynamics with the Morse Potential} \label{sec3}

Having derived the radial equation in the Dunkl-Schrödinger framework, we now investigate the quantum dynamics of a particle subject to a Morse-type interaction. The Morse potential is particularly well-suited for modeling vibrational states in diatomic molecules due to its anharmonic character. In what follows, we apply the Pekeris approximation and derive an exact analytical expression for the energy spectrum, capturing the influence of reflection symmetry through the Dunkl deformation parameters.

\subsection{The Morse Potential in the Dunkl Framework}

To proceed, we consider the Morse potential of the form:
\begin{equation}
V(r) = D \left( e^{-2\alpha(r - r_e)} - 2e^{-\alpha(r - r_e)} \right), \qquad D > 0,\ \alpha > 0,
\end{equation}
where $D$ denotes the dissociation energy, $r_e$ is the equilibrium bond length, and $\alpha$ controls the width of the potential well. This potential is attractive for large interatomic distances, reaching a minimum value $-D$ at $r = r_e$, and becomes strongly repulsive at short range.

We have to emphasize that the form of the Morse potential used here differs slightly from Eq.~\eqref{eq1} in the introduction. Although both expressions are mathematically equivalent up to a constant energy shift, the present form more directly reflects the depth of the potential well. Moreover, it facilitates algebraic simplification of the exponential terms, particularly in the application of the Pekeris approximation and the subsequent transformation of the radial equation.

To simplify the radial equation, we introduce the dimensionless variable
\begin{equation}
\chi = \frac{r - r_e}{r_e},
\end{equation}
transforming the radial equation~(\ref{eqrad}) into
\begin{equation}
\left[ \frac{d^{2}}{d\chi^{2}} - \frac{\varpi^{2} + \mu(\mu + 1)}{(\chi + 1)^{2}} 
- \frac{2Mr_e^2}{\hbar^2} D \left( e^{-2\alpha \chi} - 2e^{-\alpha \chi} \right) 
+ \frac{2Mr_e^2}{\hbar^2} E \right] \Psi(\chi) = 0.
\label{e10}
\end{equation}
In the vicinity of equilibrium, the internuclear displacement satisfies $|\chi| \ll 1$, allowing us to approximate
\begin{equation}
\frac{1}{(\chi + 1)^2} \approx 1 - 2\chi + 3\chi^2 + \cdots.
\end{equation}
Following the Pekeris prescription, we represent this expansion using exponential functions
\begin{equation}
\frac{1}{(\chi + 1)^2} \approx C_0 + C_1 e^{-\alpha \chi} + C_2 e^{-2\alpha \chi},
\label{e11}
\end{equation}
where the coefficients are given by:
\begin{equation}
C_0 = 1 - \frac{3}{\alpha} + \frac{3}{\alpha^2}, \quad
C_1 = \frac{4}{\alpha} + \frac{6}{\alpha^2}, \quad
C_2 = -\frac{1}{\alpha} + \frac{3}{\alpha^2}.
\end{equation}
Substituting Eq.~(\ref{e11}) into Eq.~(\ref{e10}) yields a modified differential equation of the form
\begin{equation}
\left[ \frac{d^{2}}{d\chi^{2}} - \eta^2 e^{-2\alpha \chi} + 2\xi^2 e^{-\alpha \chi} + W \right] \Psi(\chi) = 0,
\end{equation}
with constants :
\begin{align}
\xi^2 &= \frac{2Mr_e^2}{\hbar^2} D - \frac{C_1(\varpi^2 + \mu(\mu + 1))}{2}, \\
\eta^2 &= \frac{2Mr_e^2}{\hbar^2} D + C_2(\varpi^2 + \mu(\mu + 1)), \\
W &= (\varpi^2 + \mu(\mu + 1)) C_0 - \frac{2Mr_e^2}{\hbar^2} E.
\end{align}

\subsection{Solution of the Radial Equation}

To further simplify the equation, we introduce the transformation
\begin{equation}
\rho = e^{-\alpha \chi},
\end{equation}
which maps the domain $\chi \in (-1, \infty)$ to $\rho \in (0, e^{\alpha})$. This change of variable recasts the differential equation into the form
\begin{equation}
\left[ \rho^2 \frac{d^2}{d\rho^2} + \rho \frac{d}{d\rho} + \frac{2\xi^2}{\alpha^2} \rho 
- \frac{\eta^2}{\alpha^2} \rho^2 + \frac{W}{\alpha^2} \right] \Psi(\rho) = 0.
\label{confluente}
\end{equation}
Eq.~\eqref{confluente} is a second-order linear differential equation with regular singularities at $\rho = 0$ and an irregular singularity formally at $\rho \to \infty$. Although the physical domain of $\rho$ is bounded above by $e^{\alpha}$, it is customary to study the behavior near these singular points to guide the selection of a suitable solution form.

Near $\rho = 0$, the leading-order terms suggest that the solution behaves like a power law, $\Psi(\rho) \sim \rho^{\pm \beta}$, where $\beta$ is to be determined. At large $\rho$, the dominant contribution comes from the $\rho^2$ term with negative coefficient, indicating that normalizable solutions must decay exponentially, i.e., $\Psi(\rho) \sim e^{-\gamma \rho}$, with $\gamma > 0$.

Guided by this asymptotic structure, we propose the ansatz
\begin{equation}
\Psi(\rho) = \rho^{\beta} e^{-\gamma \rho} \Xi(\rho),
\end{equation}
where the parameters $\beta$ and $\gamma$ are defined as
\begin{equation}
\gamma = \frac{\eta}{\alpha}, \qquad \beta^2 = -\frac{W}{\alpha^2}.
\end{equation}
Substituting this ansatz into Eq.~\eqref{confluente} leads to a confluent hypergeometric-type equation for the auxiliary function $\Xi(\rho)$
\begin{equation}
\Xi(\rho) = \mathcal{C} \,\, {{}_1F_1} \left( \frac{1}{2} - \frac{\xi^2}{\gamma \alpha^2} + \beta ; 1 + 2\beta, \frac{\alpha \rho}{2\eta} \right),
\end{equation}
where {${}_1F_1(a; b; z)$} denotes the confluent hypergeometric function of the first kind { and $\mathcal{C}$ is the normalization constant.}

\subsection{Energy Spectrum and Quantum Conditions}

The requirement that the confluent hypergeometric function reduces to a polynomial imposes the quantization condition :
\begin{equation}
\frac{1}{2} - \frac{\xi^2}{\gamma \alpha^2} + \beta = -n, \qquad n = 0, 1, 2, \ldots,
\end{equation}
which ensures termination of the series and thus yields normalizable wavefunctions.

Solving for $E$, we obtain the discrete energy spectrum
\begin{equation}
E_{n,\ell,m} = \frac{\hbar^2}{2M r_e^2} \left[ \left( \mu(1 + \mu) + \varpi^2 \right) C_0 
- \alpha^2 \left( n + \frac{1}{2} - \frac{\xi^2}{\eta \alpha} \right)^2 \right],
\label{34}
\end{equation}
where $n$ denotes the radial quantum number.

A few important observations follow:
\begin{itemize}
\item The energy spectrum in Eq.~(\ref{34}) is derived under the Pekeris approximation, which becomes less accurate for large values of the angular quantum number $\ell$.
\item Bound states exist only when the argument of the square in Eq.~(\ref{34}) remains real. This yields a constraint on admissible quantum numbers:
\begin{equation}
\frac{\xi^2}{\eta \alpha} + \frac{\sqrt{(\varpi^2 + \delta(\delta + 1)) C_0}}{\alpha} - \frac{1}{2} \geq n \geq 
\frac{\xi^2}{\eta \alpha} - \frac{\sqrt{(\varpi^2 + \delta(\delta + 1)) C_0}}{\alpha} - \frac{1}{2}.
\end{equation}
\item The energy levels depend explicitly on the Dunkl deformation parameters $\mu_i$ through $\varpi^2$ and $\mu(\mu + 1)$, but remain independent of the eigenvalues of the reflection operators. Hence, the spectrum reflects deformation effects without parity splitting.
\item In the absence of deformation, i.e., when all Dunkl parameters $\mu_i$ vanish, we recover the conventional radial Schrödinger equation with the standard Morse potential. In this limit, $\varpi^2 \to \ell(\ell + 1)$, and the reflection operators no longer contribute. Thus, the Dunkl-Morse model reduces to the well-known result, confirming it as a valid generalization of the standard framework.
\end{itemize}

\section{Applications and Numerical Results} \label{sec4}

To demonstrate the physical implications of the Dunkl-Morse model, we apply our results to several diatomic molecules of experimental interest. The analysis focuses on the energy levels and their dependence on deformation parameters, quantum numbers, and molecular constants.

\subsection{Spectroscopic Parameters for Selected Molecules}

We begin by selecting three well-studied diatomic molecules—hydrogen (H\(_2\)), hydrogen chloride (HCl), and iodine (I\(_2\))—that exhibit a broad range of vibrational and rotational behaviors. The model parameters used in our computations, namely the reduced mass \(M\), equilibrium bond length \(r_e\), potential depth \(D\), and Morse width parameter \(\alpha\), are drawn from spectroscopic data and summarized in Table~\ref{tab1}.

\begin{table}[H]
\centering
\begin{tabular}{|l|l|l|l|}\hline
\hline
Molecule & $\frac{\hbar ^{2}}{2Mr_{e}^{2}}$ (cm$^{-1}$) & $D$ (cm$^{-1}$) & $\alpha$ (cm$^{-1}$) \\ \hline\hline
H$_{2}$  & 60.8296 & 38292 & 1.440 \\
HCl      & 10.5930 & 17244 & 2.380 \\
I$_{2}$  & 0.0374  & 12550 & 4954  \\ \hline
\end{tabular}
\caption{Spectroscopic parameters used for energy level calculations.}
\label{tab1}
\end{table}

These constants define the effective shape of the Morse potential for each molecule, and in combination with the Dunkl deformation parameters \(\mu_i\), they determine the behavior of the bound state energy spectrum. Notably, the large variation in the kinetic prefactor \(\hbar^2 / (2 M r_e^2)\) among the molecules leads to different energy spacing scales. For instance, H\(_2\) has the largest prefactor due to its small reduced mass and short bond length, while I\(_2\) has the smallest.

\subsection{Energy Spectra and the Role of Dunkl Parameters}

To explore the influence of the Dunkl deformation on the energy spectrum, we numerically compute energy levels \(E_{n,\ell,m}\) for each molecule using the parameters listed in Table~\ref{tab1}. We fix the angular and magnetic quantum numbers at \(\ell = m = 1\), and examine the dependence of the energy levels on the radial quantum number \(n\) for two representative values of the Dunkl parameter: \(\mu_i = -0.4\) and \(\mu_i = +0.4\). The results are presented in Tables~\ref{tab2} and \ref{tab3}.

\begin{table}[H]
  \centering
\begin{tabular}{|l|l|l|l|}\hline\hline
$n$ & H$_{2}$ & HCl & I$_{2}$ \\ \hline\hline
0  & -3.99223   & -1.97218    & -16.6989   \\ 
3  & -2.62833   & -1.31169    & -1302.44   \\ 
7  & -1.24766   & -0.639328   & -6203.1    \\ 
10 & -0.54057   & -0.29128    & -12268.8   \\ 
13 & -0.114969  & -0.0771368  & -20382.0   \\ 
15 & 0.0123821  & -0.00876624 & -26929.1   \\ 
16 & 0.0291423  & 0.00310151  & -30544.0   \\ 
17 & 0.0146258  & 0.0000909185& -34386.5   \\ 
18 & -0.0311674 & -0.017798   & -38456.6   \\ 
19 & -0.108237  & -0.0505653  & -42754.4   \\ 
20 & -0.216584  & -0.0982109  & -47279.7   \\ \hline
\end{tabular}
\caption{Energy levels \(E_{n,\ell,m}\) (eV) as a function of \(n\), for \(\ell = m = 1\) and \(\mu_i = -0.4\).}
\label{tab2}
\end{table}

\begin{table}[H]
  \centering
\begin{tabular}{|l|l|l|l|}\hline\hline
$n$ & H$_{2}$ & HCl & I$_{2}$ \\ \hline\hline
0  & -3.08793  & -1.88912    & -16.6988   \\ 
3  & -1.89144  & -1.2426     & -1302.44   \\ 
7  & -0.733998 & -0.588979   & -6203.1    \\ 
10 & -0.194321 & -0.25495    & -12268.4   \\ 
13 & 0.0638656 & -0.0548261  & -20382.0   \\ 
15 & 0.0796061 & 0.00419822  & -26929.1   \\ 
16 & 0.0405612 & 0.0113929   & -30544.0   \\ 
17 & -0.0297604& 0.00370922  & -34386.5   \\ 
18 & -0.131359 & -0.0188528  & -38456.6   \\ 
19 & -0.264234 & -0.0562931  & -42754.4   \\ 
20 & -0.428386 & -0.108612   & -47279.7   \\ \hline
\end{tabular}
\caption{Energy levels \(E_{n,\ell,m}\) (eV) as a function of \(n\), for \(\ell = m = 1\) and \(\mu_i = +0.4\).}
\label{tab3}
\end{table}

These tables highlight the sensitivity of the bound-state spectrum to the Dunkl deformation parameter. For H\(_2\) and HCl, which have relatively large kinetic prefactors \(\hbar^2 / (2 M r_e^2)\), the energy levels are closely spaced and respond strongly to changes in \(\mu_i\). In contrast, I\(_2\) exhibits much deeper and more compressed energy levels due to its small prefactor.

A noticeable feature is the asymmetry introduced by the sign of \(\mu_i\). For negative values (\(\mu_i = -0.4\)), the energy levels tend to decrease more rapidly with increasing \(n\), whereas for positive values (\(\mu_i = +0.4\)), the upper bound states are slightly elevated, indicating a softening of the effective potential. This suggests that Dunkl deformation introduces a tunable anharmonicity, which can either strengthen or weaken the confinement depending on the sign and magnitude of \(\mu_i\).

\subsection{Energy Variation with Angular Momentum}

\begin{figure}[H]
\begin{minipage}[t]{0.34\textwidth}
        \centering
        \includegraphics[width=\textwidth]{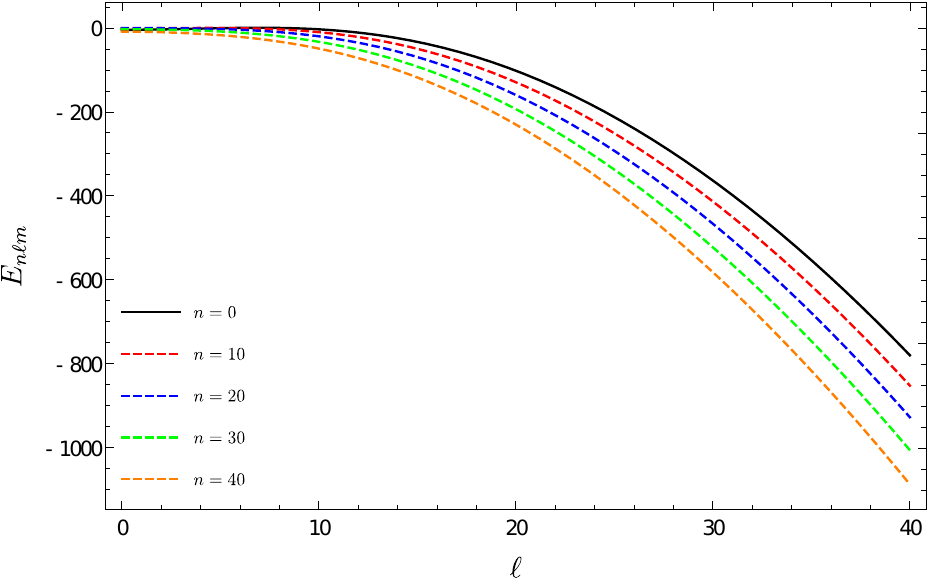}
        \subcaption{H$_{2}$}
\end{minipage}
\begin{minipage}[t]{0.34\textwidth}
        \centering
        \includegraphics[width=\textwidth]{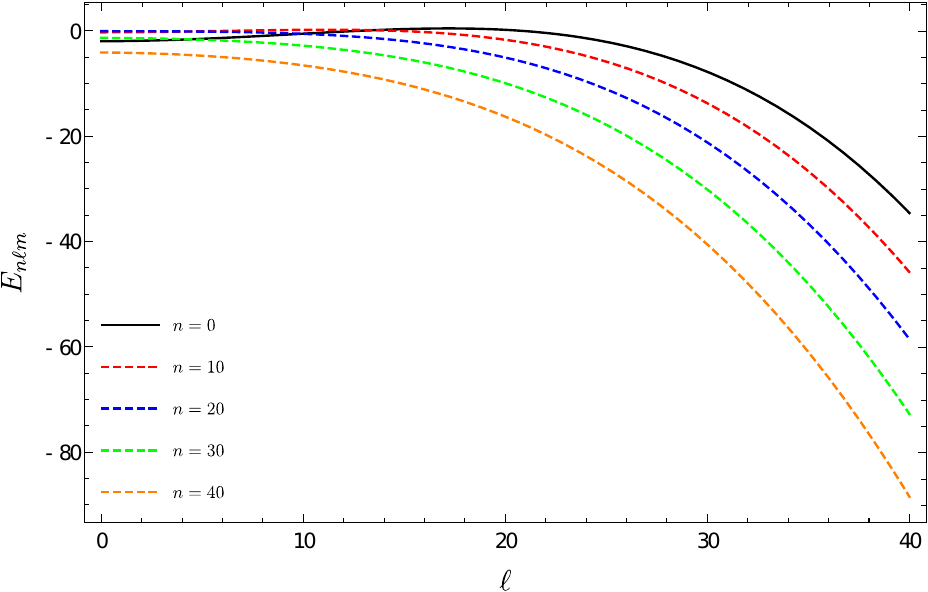}
        \subcaption{HCl}
\end{minipage}
\begin{minipage}[t]{0.34\textwidth}
        \centering
        \includegraphics[width=\textwidth]{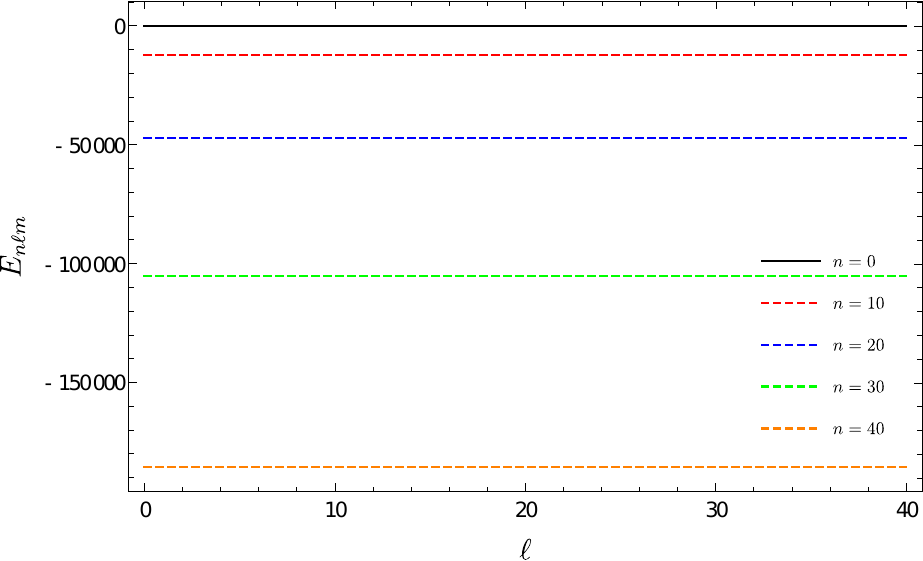}
        \subcaption{I$_{2}$}
\end{minipage}
\caption{Energy levels \(E_{n,\ell,m}\) (eV) versus angular quantum number \(\ell\), for \(n = 1\), \(m = 1\), and \(\mu_i = -0.4\).}
\label{fig:ff}
\end{figure}

\begin{figure}[H]
\begin{minipage}[t]{0.34\textwidth}
        \centering
        \includegraphics[width=\textwidth]{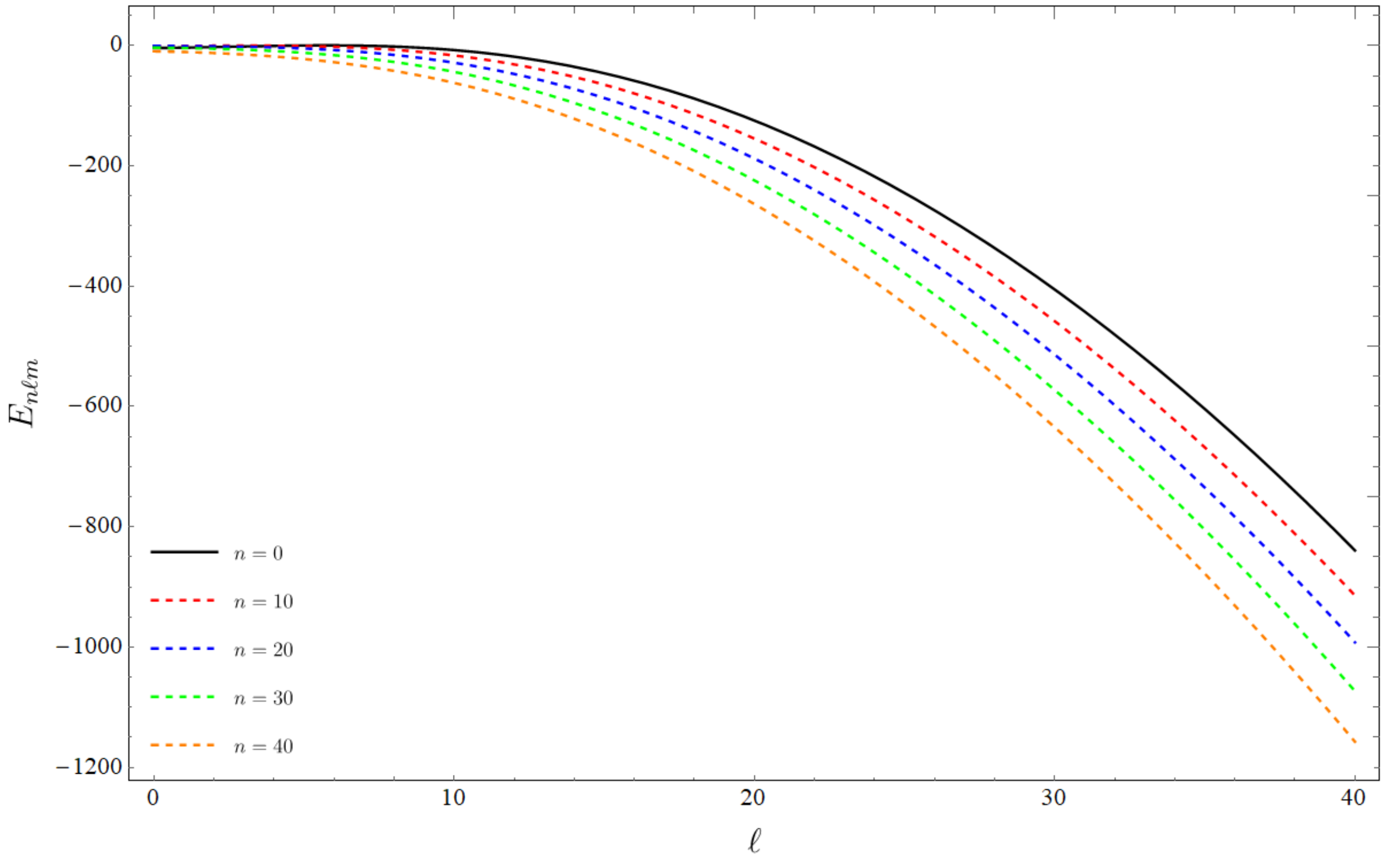}
        \subcaption{H$_{2}$}
\end{minipage}
\begin{minipage}[t]{0.34\textwidth}
        \centering
        \includegraphics[width=\textwidth]{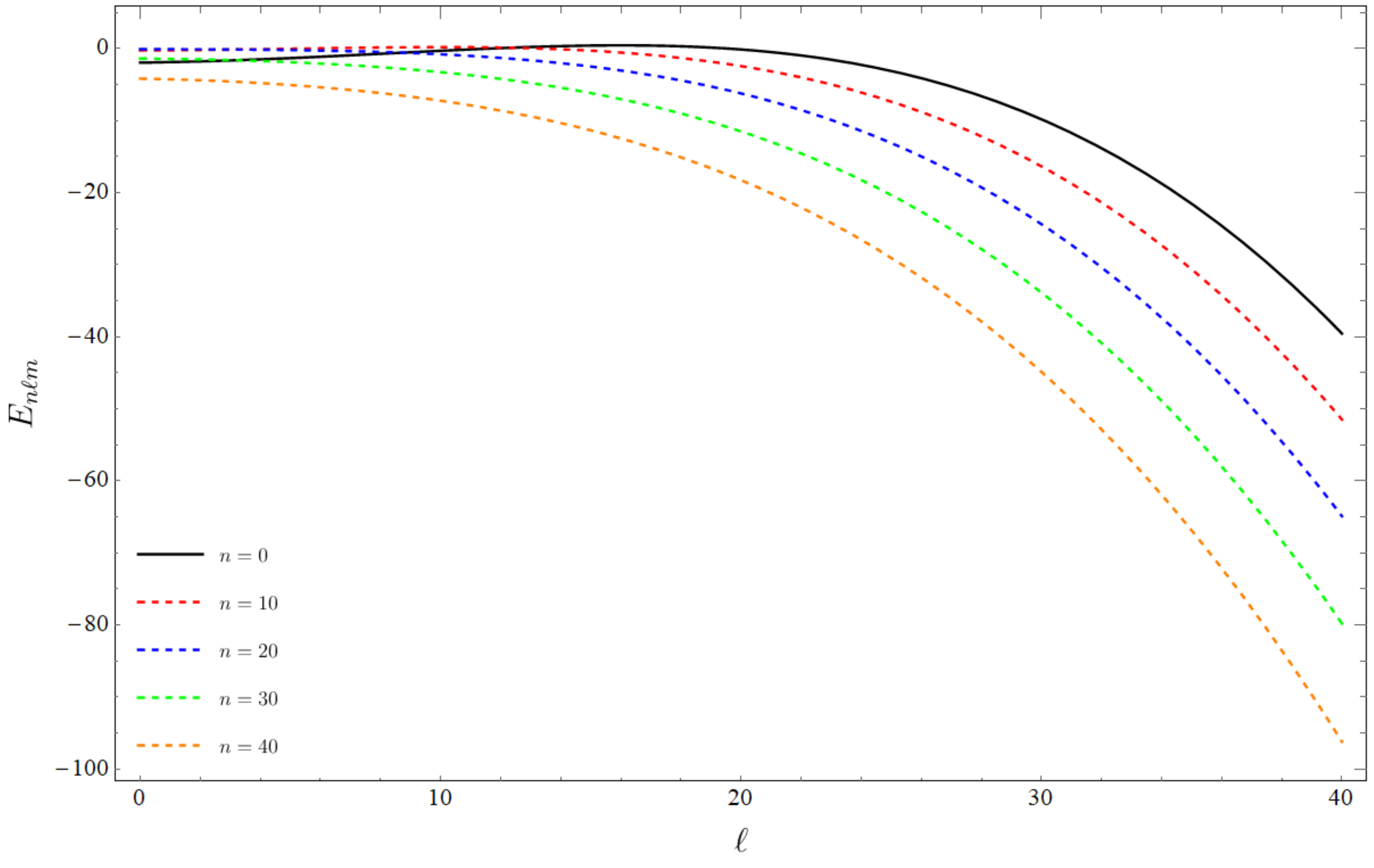}
        \subcaption{HCl}
\end{minipage}
\begin{minipage}[t]{0.34\textwidth}
        \centering
        \includegraphics[width=\textwidth]{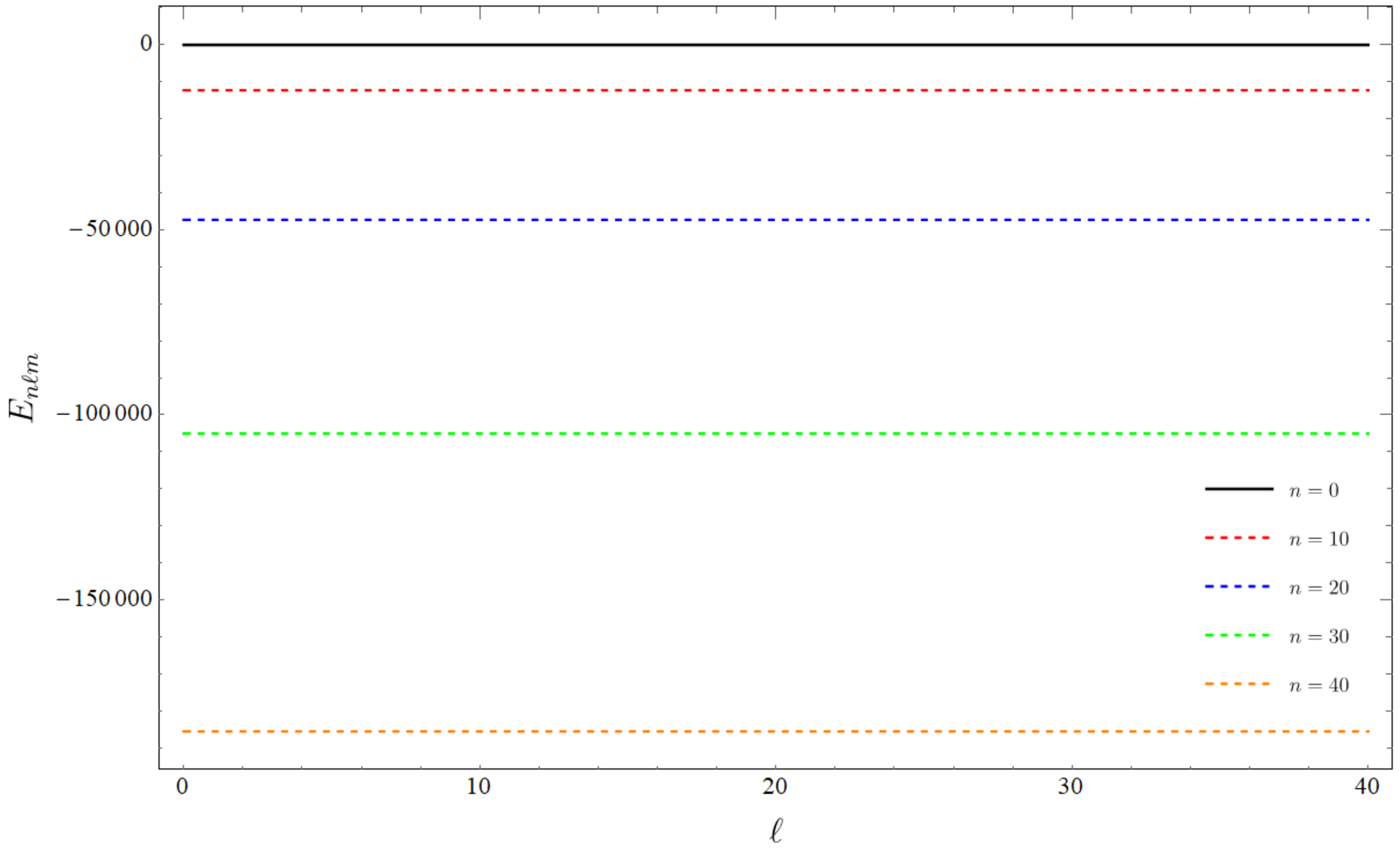}
        \subcaption{I$_{2}$}
\end{minipage}
\caption{Energy levels \(E_{n,\ell,m}\) (eV) versus angular quantum number \(\ell\), for \(n = 1\), \(m = 1\), and \(\mu_i = +0.4\).}
\label{fig:ee}
\end{figure}

To investigate the influence of angular momentum on the Dunkl-Morse spectrum, we compute the energy levels \(E_{n,\ell,m}\) as a function of the orbital quantum number \(\ell\) for a fixed radial quantum number \(n = 1\), magnetic quantum number \(m = 1\), and two representative Dunkl deformation values: \(\mu_i = -0.4\) and \(\mu_i = +0.4\). The results are visualized in Figures~\ref{fig:ff} and \ref{fig:ee} for H\(_2\), HCl, and I\(_2\).

The plots reveal several key trends. First, for all three molecules, the energy increases monotonically with \(\ell\), consistent with the growing effective repulsion due to the centrifugal term. This shift is more pronounced in lighter molecules like H\(_2\), where the kinetic prefactor is larger. Second, the Dunkl parameter modulates this trend: for \(\mu_i < 0\), the potential is effectively deeper, leading to more negative energies; for \(\mu_i > 0\), the opposite occurs, and the spectrum shifts upward.

It is important to note that the results presented in Figures~\ref{fig:ff} and \ref{fig:ee} extend to relatively large values of \(\ell\). While the Pekeris approximation offers an analytically tractable means of incorporating the centrifugal term, its accuracy diminishes for high angular momentum. Therefore, the results at large \(\ell\) should be interpreted with caution and regarded as qualitative indicators of spectral trends rather than precise quantitative predictions.

Despite this limitation, the visualizations clearly demonstrate how angular momentum and symmetry deformation interact in shaping the energy landscape of the Dunkl-Morse system. The framework provides a flexible and physically transparent means of exploring symmetry-modified vibrational dynamics in diatomic systems.

\section{Thermal Properties} \label{sec5}

The Morse potential has played a pivotal role in the development of molecular spectroscopy and quantum thermodynamics, tracing back to Dunham’s pioneering work on vibrational spectra and thermodynamic behavior in diatomic molecules~\cite{r11}. In this section, we investigate the thermodynamic behavior of the Dunkl-Morse system using the analytical energy expression derived in Eq.~\eqref{34}. Based on this spectrum, we compute key thermal functions—partition function, free energy, internal energy, entropy, and specific heat capacity—and analyze their dependence on both temperature and the Dunkl deformation parameter.

\subsection{Partition Function and Thermodynamic Formalism}

The vibrational partition function is defined as
\begin{equation}
Z(\beta) = \sum_{n=0}^{\lambda} e^{-\beta E_n}, \qquad \beta = \frac{1}{k_B T},
\end{equation}
where \(k_B\) is the Boltzmann constant, \(T\) is the absolute temperature, and \(\lambda\) is the upper limit of the vibrational quantum number, determined from the constraint \(\frac{dE_n}{dn} = 0\).

To evaluate the summation analytically, we adopt a lower-order approximation of the Poisson summation formula. The resulting expression for the partition function is
\begin{align}
Z(\beta) &= \frac{1}{2} \left[ e^{-\beta P (Q - (\alpha H)^2)} - e^{-\beta P (Q - \alpha^2 (\lambda + 1 + H)^2)} \right] \nonumber \\
&\quad + \frac{\sqrt{\pi} e^{-\beta P Q} \left( \text{erfi} \left( \alpha \sqrt{\beta P} (H + \lambda + 1) \right) - \text{erfi} \left( \alpha H \sqrt{\beta P} \right) \right)}{2 \alpha \sqrt{\beta P}},
\label{x1}
\end{align}
with the following parameter definitions
\begin{equation}
P = \frac{\hbar^2}{2 M r_e^2}, \quad Q = \mu (\mu + 1) C_0, \quad H = \frac{1}{2} - \frac{\xi_1^2}{\eta_1 \alpha}, \quad \lambda = -H,
\end{equation}
\begin{equation}
\xi_1 = \sqrt{ \frac{2 M r_e^2}{\hbar^2} D - \frac{C_1 \mu(\mu + 1)}{2} }, \quad 
\eta_1 = \sqrt{ \frac{2 M r_e^2}{\hbar^2} D + C_2 \mu(\mu + 1) }.
\end{equation}

The remaining thermodynamic functions are then calculated via the standard relations
\begin{equation}
F(\beta) = -\frac{1}{\beta} \ln Z(\beta), \qquad 
U(\beta) = -\frac{\partial}{\partial \beta} \ln Z(\beta),
\end{equation}
\begin{equation}
S(\beta) = k_B \left[ \ln Z(\beta) - \beta \frac{\partial \ln Z(\beta)}{\partial \beta} \right], \qquad 
C_v(\beta) = k_B \beta^2 \frac{\partial^2 \ln Z(\beta)}{\partial \beta^2}.
\end{equation}

\subsection{Thermal Behavior of Diatomic Molecules}

\begin{figure}[H]
\centering
\begin{minipage}[t]{0.32\textwidth}
    \includegraphics[width=\textwidth]{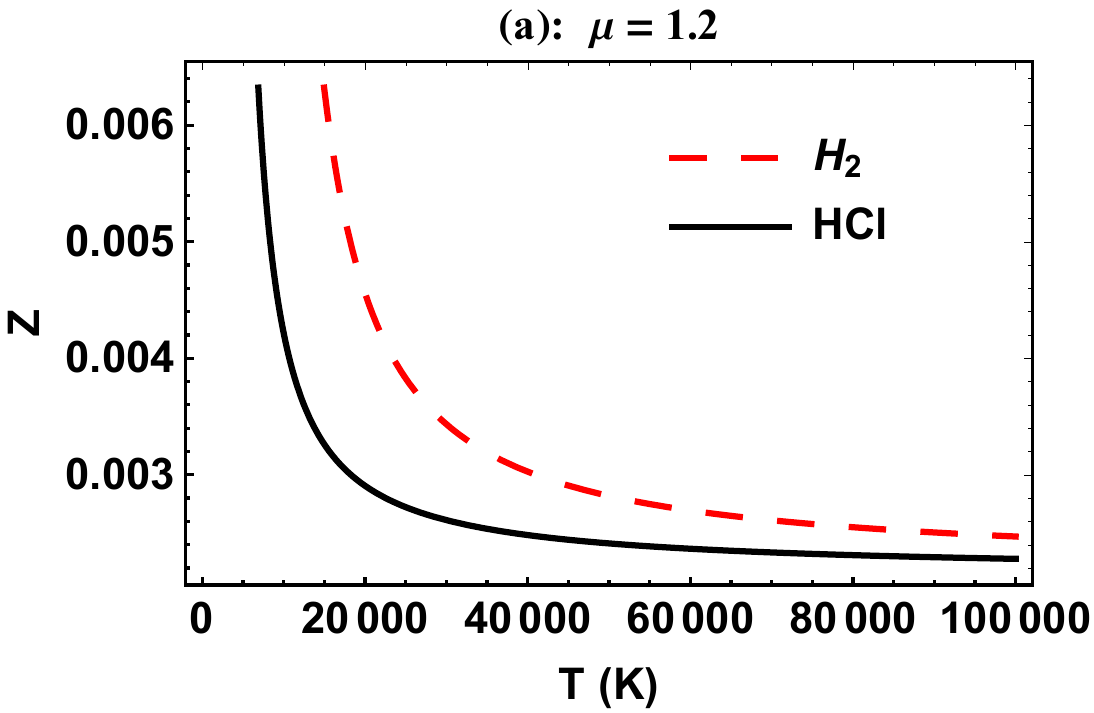}
    \subcaption*{(a) Partition function \(Z(T)\)}
\end{minipage}
\begin{minipage}[t]{0.32\textwidth}
    \includegraphics[width=\textwidth]{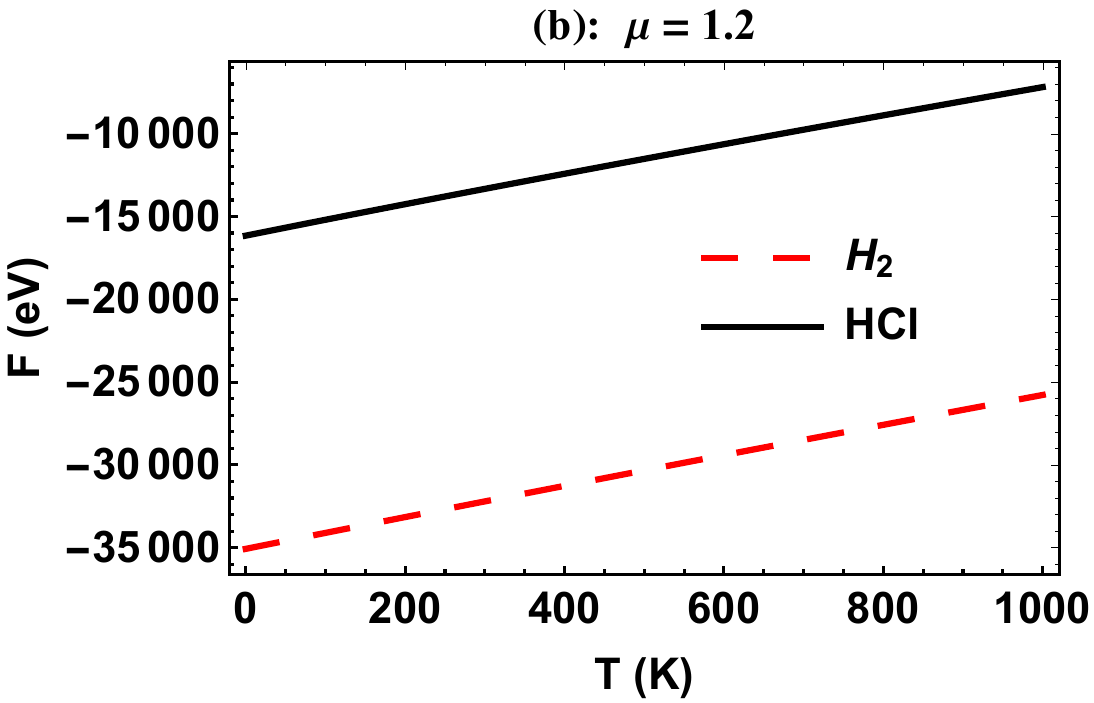}
    \subcaption*{(b) Free energy \(F(T)\)}
\end{minipage}
\begin{minipage}[t]{0.32\textwidth}
    \includegraphics[width=\textwidth]{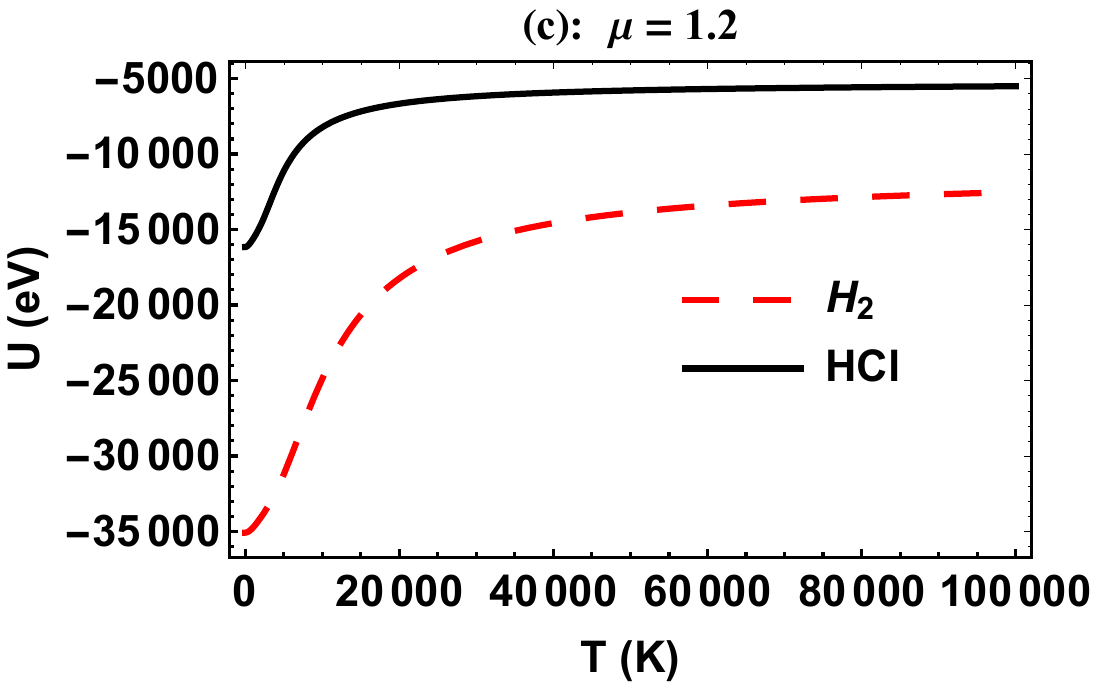}
    \subcaption*{(c) Internal energy \(U(T)\)}
\end{minipage}

\begin{minipage}[t]{0.32\textwidth}
    \includegraphics[width=\textwidth]{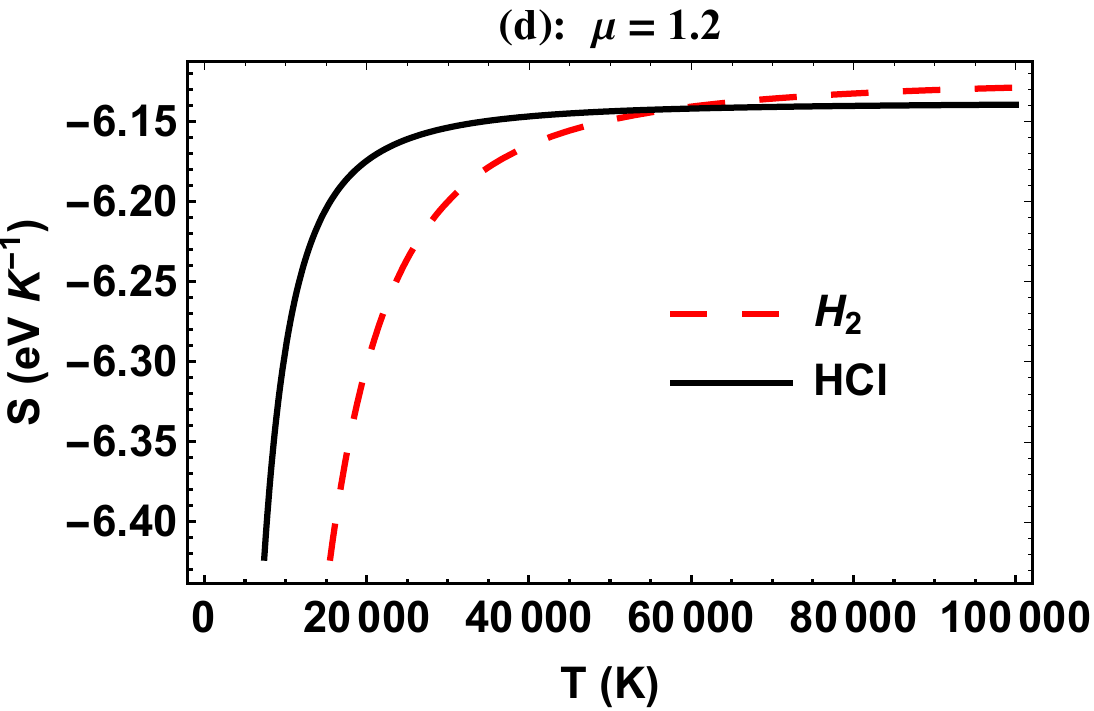}
    \subcaption*{(d) Entropy \(S(T)\)}
\end{minipage}
\begin{minipage}[t]{0.32\textwidth}
    \includegraphics[width=\textwidth]{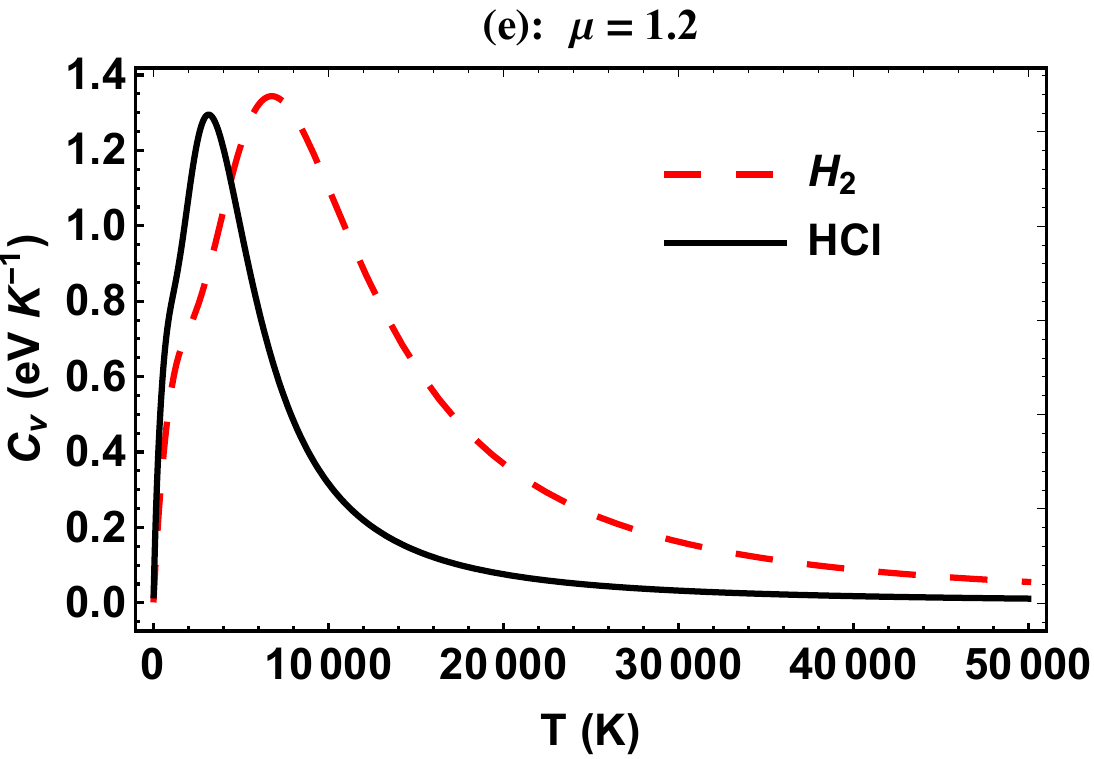}
    \subcaption*{(e) Specific heat \(C_v(T)\)}
\end{minipage}
\caption{Thermodynamic functions of H\(_2\) and HCl for \(\mu = 1.2\).}
\label{fig:xx}
\end{figure}

Figure~\ref{fig:xx} presents the thermodynamic functions of two representative diatomic molecules, H\(_2\) and HCl, for a fixed Dunkl parameter \(\mu = 1.2\).

The partition function (a) decreases and saturates at high temperatures, with H\(_2\) displaying a higher magnitude due to its broader vibrational spectrum. In (b), the free energy increases linearly with temperature, more steeply for HCl. Internal energy (c) and entropy (d) grow monotonically and eventually stabilize, reflecting the finite number of accessible energy states. The specific heat (e) peaks at intermediate temperatures, consistent with the Dulong–Petit behavior, and declines as thermal saturation is reached.

\subsection{Effect of Dunkl Deformation on Thermal Response}

\begin{figure}[H]
\centering
\begin{minipage}[t]{0.32\textwidth}
    \includegraphics[width=\textwidth]{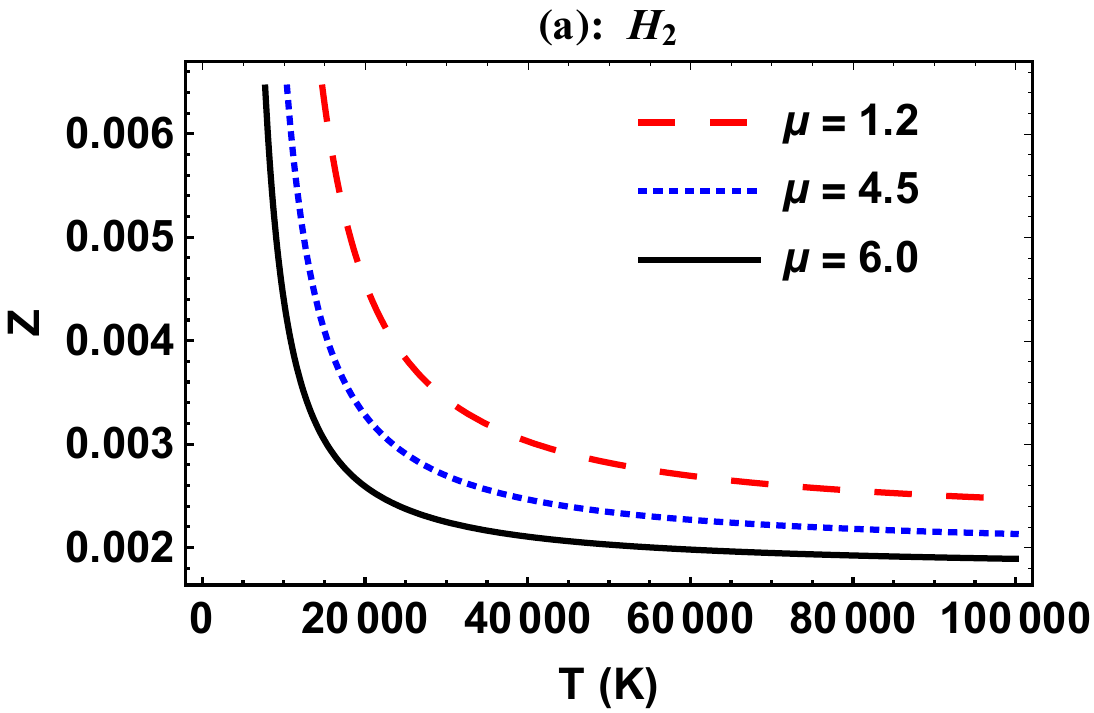}
    \subcaption*{(a) Partition function \(Z(T)\)}
\end{minipage}
\begin{minipage}[t]{0.32\textwidth}
    \includegraphics[width=\textwidth]{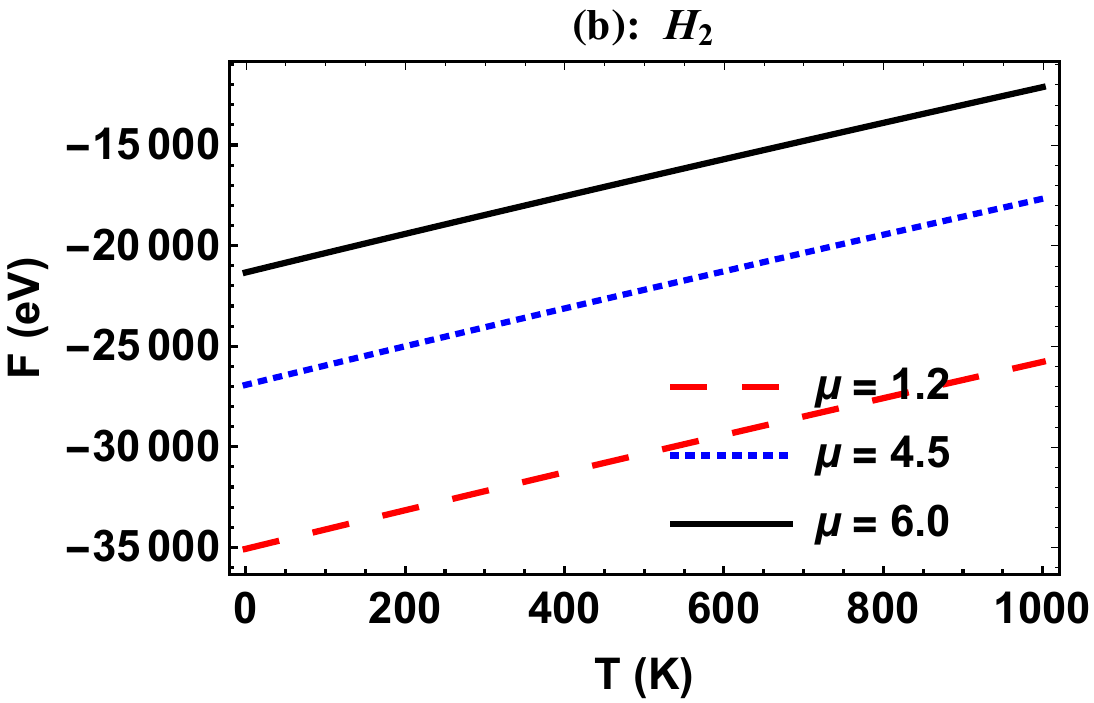}
    \subcaption*{(b) Free energy \(F(T)\)}
\end{minipage}
\begin{minipage}[t]{0.32\textwidth}
    \includegraphics[width=\textwidth]{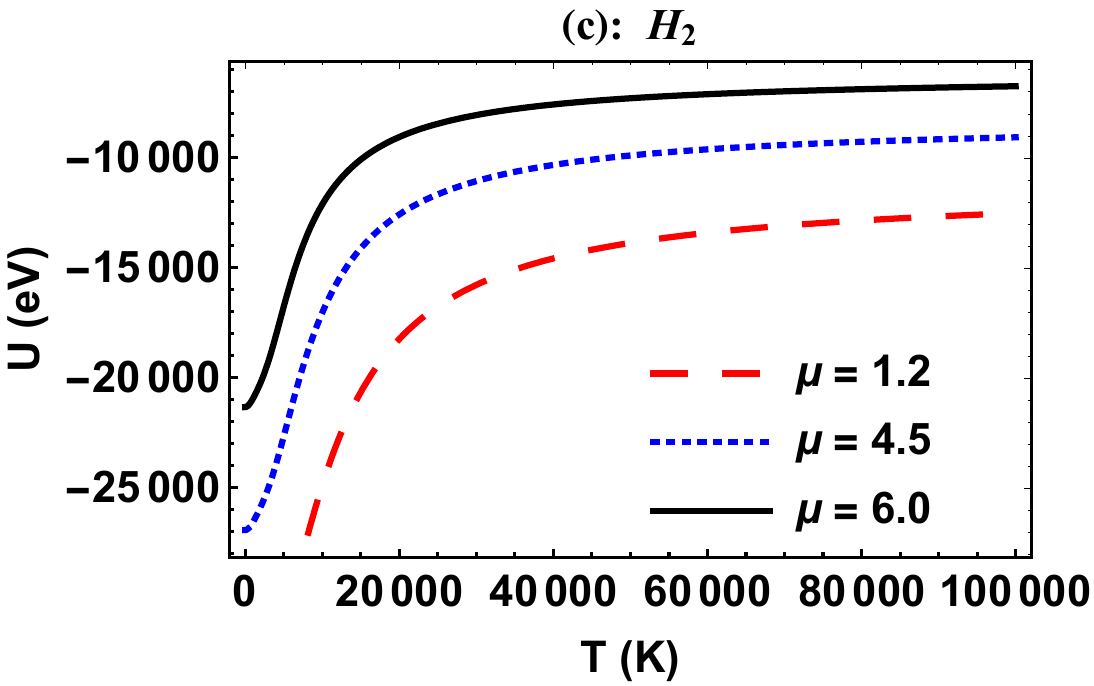}
    \subcaption*{(c) Internal energy \(U(T)\)}
\end{minipage}

\begin{minipage}[t]{0.32\textwidth}
    \includegraphics[width=\textwidth]{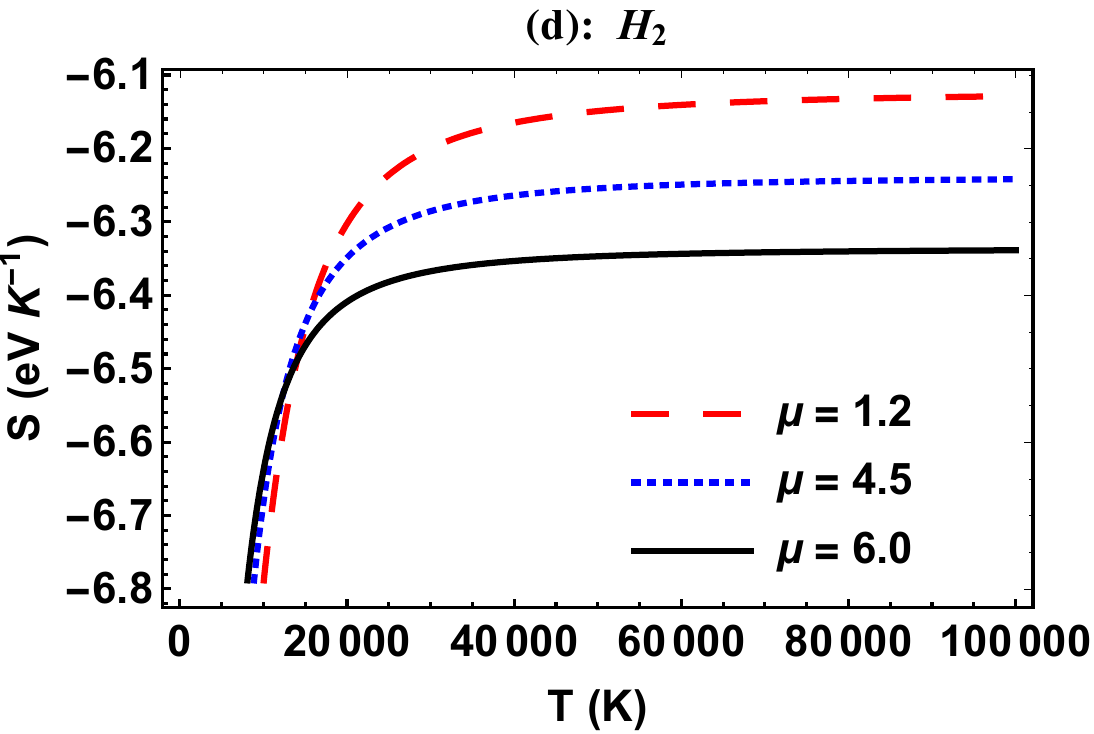}
    \subcaption*{(d) Entropy \(S(T)\)}
\end{minipage}
\begin{minipage}[t]{0.32\textwidth}
    \includegraphics[width=\textwidth]{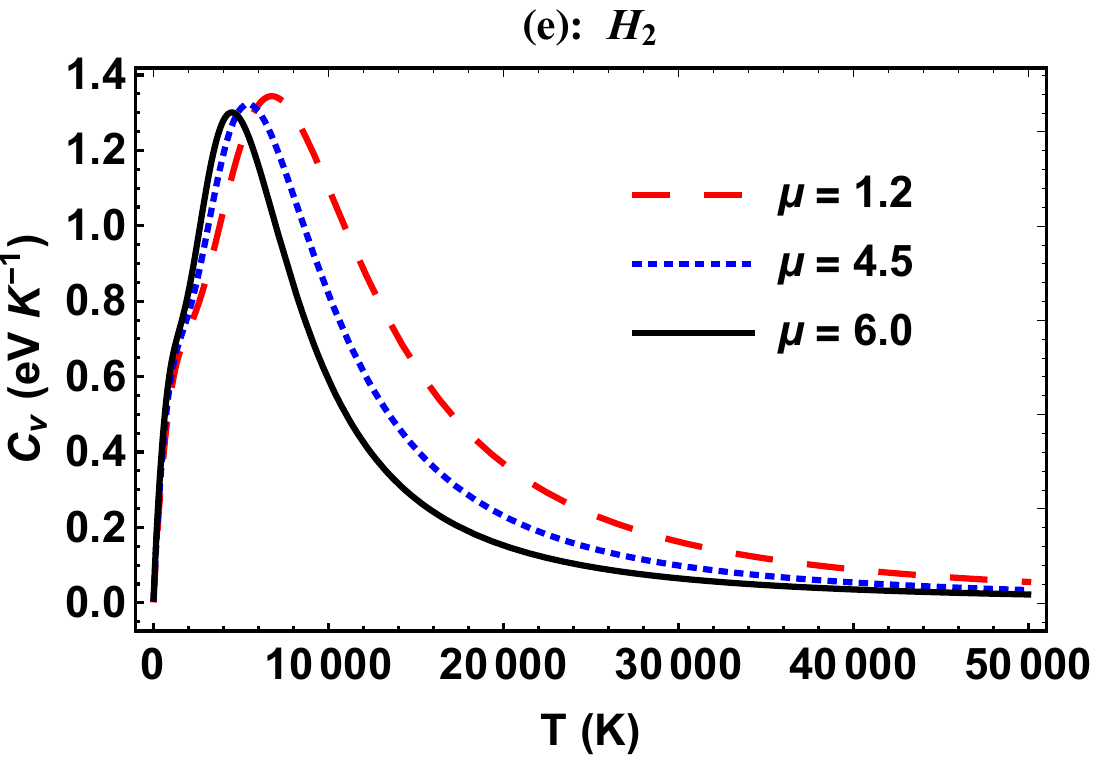}
    \subcaption*{(e) Specific heat \(C_v(T)\)}
\end{minipage}
\caption{Thermal functions of H\(_2\) under different Dunkl deformation parameters \(\mu = 1.2, 4.5, 6.0\).}
\label{fig:yy}
\end{figure}

Figure~\ref{fig:yy} highlights the role of Dunkl deformation in shaping the thermodynamic behavior of H\(_2\), by varying \(\mu\) while keeping molecular parameters fixed.

As \(\mu\) increases, the partition function (a) decreases due to a narrower vibrational bandwidth. The free energy (b) rises sharply with increasing \(\mu\), and the internal energy (c) becomes larger, reflecting stronger confinement. The entropy (d) exhibits an inverse trend—decreasing with higher \(\mu\), indicating fewer thermally accessible states. The specific heat (e) retains its peak behavior, but the peak shifts to lower values and flattens as \(\mu\) increases.

These results confirm that the Dunkl deformation not only modifies the spectral properties of the system but also leaves a significant imprint on its thermodynamic profile. The framework thus offers a tunable bridge between quantum mechanical deformation and molecular statistical mechanics.

\section{Conclusion} \label{sec6}

In this work, we investigated the bound-state solutions and thermodynamic properties of the conventional Morse potential within the framework of Dunkl quantum mechanics. By introducing Dunkl derivatives—operators that encode reflection symmetry—into the radial Schrödinger equation, we derived a modified wave equation that captures the interplay between molecular anharmonicity and discrete parity-altering effects.

Applying the Pekeris approximation to handle the centrifugal term, we successfully reduced the problem to a solvable form and obtained exact analytical expressions for both the energy eigenvalues and eigenfunctions. These results provide a comprehensive description of the spectral structure, demonstrating how Dunkl deformation parameters modify level spacing and spectral density. The derivation of a closed-form partition function, based on the exact spectrum, enabled us to compute key thermal quantities—including free energy, internal energy, entropy, and specific heat—with explicit dependence on both the temperature and Dunkl parameters.

The model was applied to several diatomic molecules, revealing that the Dunkl formalism introduces a tunable and physically interpretable deformation to the standard quantum behavior, especially in vibrational spectra and thermal response. This framework not only enriches the analytical treatment of molecular systems but also offers a novel pathway to incorporate symmetry-induced modifications in quantum models.

These findings highlight the technical power and physical versatility of combining conventional potentials with Dunkl operators. Future research may extend this approach to time-dependent or relativistic settings, explore interactions with external fields, or generalize it to multi-particle and higher-dimensional systems where reflection symmetries and deformations play a central role.


\section*{Acknowledgments}
{The authors would like to thank the anonymous referees for their valuable comments and suggestions, which helped improve the clarity and quality of this work.}B. C. L. is grateful to Excellence Project PřF UHK 2211/2023-2024 for the financial support.

\section*{Data Availability Statements}

The authors declare that the data supporting the findings of this study are available within the article.

\end{document}